\pdfoutput=1

\documentclass[11pt]{article}
\usepackage{authblk}

\usepackage[preprint]{EMNLP2023}

\usepackage{times}
\usepackage{latexsym}

\usepackage{graphics}
\usepackage{CJK}
\usepackage{tabularx}
\usepackage{bm}
\usepackage{amssymb}
\usepackage{booktabs}
\usepackage{url}
\usepackage{multicol}
\usepackage{color}
\usepackage{arydshln}
\usepackage{multirow}
\usepackage{amsmath} 
\usepackage{amsfonts}
\usepackage{subfigure}
\usepackage{amstext}
\usepackage{tikz}
\usepackage{pgfplots}
\usepackage{hyperref}
\usepackage{comment}
\usepackage{bbding}
\usepackage{multirow}
\usepackage{pgfplots}
\pgfplotsset{compat=1.18}

\usepackage[justification=justified,skip=6pt]{caption}

\usepackage{adjustbox}

\usepackage[T1]{fontenc}

\usepackage[utf8]{inputenc}

\usepackage{microtype}

\usepackage{inconsolata}

\title{Safely Learning with Private Data: \\A Federated Learning Framework for Large Language Model}

\author{
  \textbf{JiaYing Zheng}\textsuperscript{1}, 
  \textbf{HaiNan Zhang}\textsuperscript{1 $*$}\thanks{\textit{$*$ Corresponding author.}}, 
  \textbf{LingXiang Wang}\textsuperscript{1}, 
  \textbf{WangJie Qiu}\textsuperscript{1},\\ 
  \textbf{HongWei Zheng}\textsuperscript{2}, 
  \textbf{ZhiMing Zheng}\textsuperscript{}
}

\affil[1]{Beijing Advanced Innovation Center for Future Blockchain and Privacy Computing, 
\authorcr Institute of Artificial Intelligence, Beihang University, China}
\affil[2]{Beijing Academy of Blockchain and Edge Computing, China}
\affil[ ]{\texttt{\{jiayingzheng, zhanghainan,  wanglingxiang\}@buaa.edu.cn}}

\makeatletter
\def\thanks#1{\protected@xdef\@thanks{\@thanks
        \protect\footnotetext{#1}}}
\makeatother

\begin{document}

\maketitle

\begin{abstract}
 
Private data, being larger and quality-higher than public data, can greatly improve large language models (LLM). However, due to privacy concerns, this data is often dispersed in multiple silos, making its secure utilization for LLM training a challenge. 
Federated learning (FL) is an ideal solution for training models with distributed private data, but traditional frameworks like FedAvg are unsuitable for LLM due to their high computational demands on clients. An alternative, split learning, offloads most training parameters to the server while training embedding and output layers locally, making it more suitable for LLM. Nonetheless, it faces significant challenges in security and efficiency. Firstly, the gradients of embeddings are prone to attacks, leading to potential reverse engineering of private data. Furthermore, the server's limitation of handle only one client's training request at a time hinders parallel training, severely impacting training efficiency.
In this paper, we propose a Federated Learning framework for LLM, named FL-GLM, which prevents data leakage caused by both server-side and peer-client attacks while improving training efficiency. 
Specifically, we first place the input block and output block on local client to prevent embedding gradient attacks from server. Secondly, we employ key-encryption during client-server communication to prevent reverse engineering attacks from peer-clients. Lastly, we employ optimization methods like client-batching or server-hierarchical, adopting different acceleration methods based on the actual computational capabilities of the server.
Experimental results on NLU and generation tasks demonstrate that FL-GLM achieves comparable metrics to centralized chatGLM model, validating the effectiveness of our federated learning framework.

\end{abstract}

\section{Introduction}

Existing large language models (LLM) have achieved astonishing results by utilizing vast amounts of public data and massive parameters. In comparison to public data, private data holds advantages in both quantity and quality, because private datasets typically encompass more comprehensive and detailed information about individuals or organizations, and the data production process is more rigorous. Therefore, private data can undoubtedly further enhance the performance of LLM.

\begin{figure}[!t]
    \centering
    \includegraphics[width=1.0\linewidth]{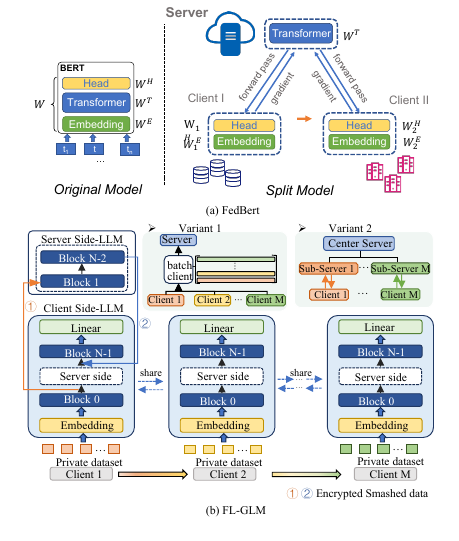}
    \caption{Model architecture of FedBert and FL-GLM. }
    \label{fig:compare}
\end{figure}

However, private data is often stored in isolated data silos. For example, mobile users' data is kept locally, involving a significant amount of personal privacy. Considering privacy and security, LLM cannot store private data in a centralized manner for training. Hence, securely leveraging private data for language model training remains a challenging problem.
An ideal solution is to utilize the Federated Learning (FL)~\citep{li2020review} framework, which allows data to be retained on the user device for local training and only passes the model parameters to server for model aggregation. 
This approach achieves the goal of keeping the data stationary while making the model updates. By using FL for LLM training, data privacy can be preserved, and the performance of LLM can be further improved.

Unfortunately, traditional FL frameworks, such as FedAvg~\citep{stremmel2021pretraining} and FedProx\citep{li2020federated}, are not suitable for LLMs because they require each client to have sufficient computational resources to train the entire LLM. 
As an alternative method, transformer with split learning, represented by FedBERT~\citep{tian2022fedbert} in Figure~\ref{fig:compare}(a), focuses most of the parameters on the server while continuously training the embedding layer and output layer on the local client, making it more suitable for LLM. The process involves the client using the embedding layer for data input and forwarding it to the server, which then computes and returns the output states. The client calculates loss, sends gradients back for server updates, and receives updated gradients for the embedding layer. However, this method presents security risks. Embedding gradients are vulnerable to attacks~\citep{yaldiz2023secure}, potentially allowing attackers to reconstruct private data through beam search~\citep{gupta2022recovering} or reverse-engineered~\citep{asnani2023reverse}. Additionally, since the server processes one client at a time, it hinders parallel training and reduces efficiency.

In this paper, we propose a FL framework called FL-GLM for LLM, as shown in Figure~\ref{fig:compare}(b). We partition the transformers of the chatGLM into three parts: input and output blocks are stored on the client (shared with peers), while the remaining large parameters are kept on the server. During the training process, the client first performs forward propagation on the input data to obtain hidden states. Then, these hidden states are encrypted with a secure key and sent to the server. Subsequently, the server, either in a client-batch or server-layered approach, receives more client hidden states at a training time and executes forward propagation to send output hidden states back to each client.

It is clear that our FL-GLM framework can effectively prevent data leakage attacks from both servers and peer-clients while enhancing training efficiency. Clients and servers jointly own and utilize the entire model, with certain input and output blocks placed on local clients to thwart embedding gradient attacks from the server. Although sharing input and output blocks between all clients can improve results, interception by peers poses risks, which can be resolved through key encryption during client-server communication. To overcome server capacity limitations, we propose various training acceleration methods. For clusters with multiple machines and GPUs, a hierarchical server architecture initializes sub-servers for parallel client training, with central server aggregating and distributing models. With single machines and multiple GPUs, the client-batch method concatenates client information for training, enabling parallel execution and enhanced efficiency compared to traditional serial execution in split learning.

Experimental results on NLU and generation tasks demonstrate that FL-GLM achieves performance comparable to centralized chatGLM-6B models, validating the effectiveness of our framework. Further analysis of training costs indicates that our client-batch and server-hierarchical mechanisms can save more than 48\% of training time.
Our code is available at: \href{ https://github.com/TAP-LLM/SplitFedLLM}{https://github.com/TAP-LLM/SplitFedLLM.}

The innovations in this paper are as follows:
\begin{itemize}
   
    \item To the best of our knowledge, we are the first to design a federated learning framework specifically tailored for LLMs. Starting from user privacy concerns and considering the computational demands of LLMs, we improve split learning to adapt to LLMs, and develop a reasonable, effective, and secure federated LLM framework.
    \item We propose client-batch and server-hierarchical acceleration optimization methods based on the server's computational capacity to address the issue of low training efficiency in split learning.
    \item  Experimental results on SuperGLUE and abstractive summarization datasets demonstrate that the proposed FL-GLM model can obtain comparable performance to centralized chatGLM models, validating the effectiveness of our FL framework.
\end{itemize}

\section{Related Work}
\subsection{Federated Learning in LM}
Federated Learning (FL) has emerged as a promising approach to train language models (LM) in a decentralized manner while respecting user privacy and data safety. 
Federated Averaging (FedAvg)~\citep{mcmahan2017communication} is a popular federated optimization algorithm used in language models~\citep{hard2018federated,chen2019federated,stremmel2021pretraining}. In FedAvg, each client trains its model on locally stored data and communicates updates to the server. The server then performs weighted aggregation of these updates to create a new global model. To reduce local training rounds and accelerate the learning process, ~\citep{stremmel2021pretraining} proposes to utilize the pre-trained global models on FedAvg. ~\citep{ji2019learning} proposes Attentive Federated Aggregation (FedAtt) and applies a layer-wise soft attention mechanism to the trained parameters of the neural network model.
Previous works~\citep{jalalirad2019simple, thakkar2020understanding} have integrated DP mechanisms into FedAvg and FedAtt, respectively.

Split learning, represented by SplitFed~\citep{thapa2022splitfed}, has emerged as a distributed and collaborative training approach to enable efficient training on resource-constrained devices~\citep{abedi2020fedsl,abuadbba2020can,rahman2020internet,matsubara2020neural}, such as mobile devices or small clients without GPU resources.
To address sequential data training in language models, FedBERT~\citep{tian2022fedbert} introduces a novel federated learning framework. It splits language model pre-training, easing limited computing resources on client devices. FedBERT segments the BERT model into Embedding, Transformer, and Output layers. It trains the Transformer layer on a powerful server, while less demanding layers (Embedding and Output) train on client devices. However, this setup incurs high communication costs and risks data leakage via embedding gradient attacks.

\subsection{Attacks and Defenses}

In federated learning, various eavesdroppers threaten client privacy, including servers attempting data recovery and peer-clients intercepting data sent to servers.In NLP, attacks from embedding gradients can easily recover users's private data. ~\citep{gupta2022recovering} proposes to infer which words the client used by observing the non-zero values in embedding gradients. They then use beam search and resort to arrange these words, thereby reconstructing private data. 
To counter this, they recommend freezing embedding layers during training. ~\citep{zhu2019deep} briefly mentions defending by adding differentially private noise or setting small gradients to zero (gradient clipping). ~\citep{huang2020texthide} propose MixUp data augmentation on the BERT model's [CLS] token. ~\citep{yaldiz2023secure} suggest server-side cosine similarity checks on client-uploaded weights to filter out malicious clients. However, these defenses often reduce model accuracy~\citep{yu2021differentially, li2021large}.

In order to retain the model structure and minimize the performance loss caused by model changes, we propose to move some head layers to the client and use a key-encryption mechanism to protect data privacy during client-server communication. This not only prevents gradient attacks from the server but also prevents information eavesdropping from peers.

\section{Model}
In this section, we provide the details of the FL-GLM framework, as shown in Figure ~\ref{fig:compare}(b). FL-GLM consists of three parts: model split, encrypted transmission, and parallel acceleration. Firstly, we split LLM into three parts, saving the first block 0 and the last block N-1 on the local client and placing the remaining parameters on the server. Then, the smashed data is encrypted using keys during client-server transmission. Finally, the server employs either client-batch or hierarchical-server methods to achieve parallel acceleration.

\begin{figure*}[!t]
    \centering
    \includegraphics[width=1.0\linewidth]{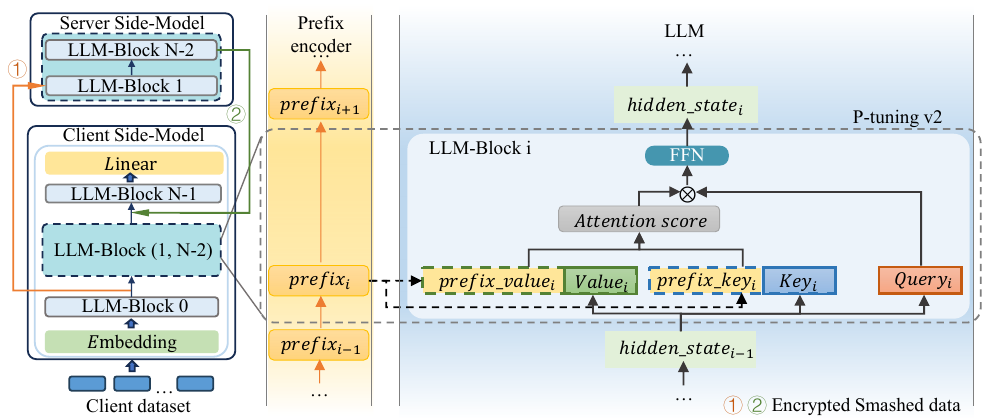}
    \caption{Model Split with p-tuning v2 fine-tuning by training a prefix encoder to adjust LLM-Block outputs. 
    }
    \label{fig:framework}
\end{figure*}

\subsection{Model Split}

For protecting data privacy, the FL-GLM framework splits LLM into three parts for deployment. 
Take ChatGLM as an example, the complete ChatGLM model contains the embedding layer, 28 ChatGLM blocks and the final linear layer, the client side contains the embedding layer, the 0th ChatGLM, the 27th ChatGLM block and the final linear layer, and the 1st to 26th ChatGLM are deployed on the server side, each block is a transformer structure. 
During forward operations, the client-side model processes private data to generate smashed data, which is then sent to the server-side model for computation. Encrypting the smashed data ensures its security.
Given the input data $x=\left\{x_1,\dots,x_L \right\}$ and the next output $y$, the smashed data $h_{0}$ of the client is defined as:
\begin{equation*}
   h_{0}=Block_0\left(Embedding\left (x \right ) \right),  
\end{equation*}
where $Block_0$ is the $0^{th}$ block of LLM, and $Embedding$ is the embedding layer of LLM.

The server-side model contains the $1^{th}$ to the $\text{N-2}^{th}$ blocks of LLM, denoted as $Block_{(1,\text{N-2})}$, which takes the received smashed data $h_{0}$ as input, and the hidden state $h_{\text{N-2}}$ as output:
\begin{equation*}
  h_{\text{N-2}} = Block_{(1,\text{N-2})}(h_{0}). 
\end{equation*}
Then the server send the output $h_{\text{N-2}}$ back to client.

After the last block N-1 and the linear layer operation on client, the prediction result $y^{'}$ is output and the cross-entropy loss $\mathcal{L}$ is calculated:
\begin{align*}
  y^{'} &= Linear\left(Block_{\text{N-1}} \left(h_{\text{N-2}}\right) \right), \\
  \mathcal{L} &=Cross\_Entropy\left (y^{'}, y \right),
\end{align*}
where $Block_{\text{N-1}}$ is the $\text{N-1}^{th}$ block of LLM and $Linear$ is the linear layer of LLM.
During the whole computation process, the data and data labels are kept in the client to avoid data privacy leakage.

It's important to note that the LLM-Block is constructed from a transformer layer comprising multi-head self-attention mechanisms and a forward network (FFN). With the stacking of LLM-Blocks, large pre-trained models have an extremely high number of parameters, making fine-tuning computationally intensive. To fine-tune large models with limited computational resources, efficient techniques such as p-tuning v2~\citep{liu2021p} can be employed, as depicted in Figure~\ref{fig:framework}. The FL-GLM framework supports the p-tuning v2 method, wherein all original model parameters are frozen, and the prefix encoder is trained to splice the prefix\_key and prefix\_value with the key and value of the original model, adjusting the output of each LLM-Block. Further details see in Appendix~\ref{sec:model-details}.

\subsection{Encrypted Transmission}
Since the data features need to flow between the client and the server after the model split, the FL-GLM framework uses a key encryption strategy to complete the encrypted transmission of data.
The RSA algorithm generates a pair of public and private keys by factorizing a very large integer. The message is encrypted with the public key and can only be decrypted by the receiver who has the corresponding private key. The RSA key generation process is as follows:

1) Select two large prime numbers, usually denoted as $p$ and $q$.

2) Calculate their product $n=pq$. $n$ will be used as the common modulus.

3)  Compute the Euler's totient function $\phi \left(n \right) = \left(p-1 \right)\left(q-1 \right)$. For a prime number $p$, there are $p-1$ numbers that are coprime with $p$; similarly, for a prime number $q$, there are $q-1$ numbers that are coprime with $q$. Since $p$ and $q$ are coprime with each other, the Euler's totient values for $p$ and $q$ can be multiplied directly.

4) Choose an integer $e$, called the public key exponent, satisfying $1 < e < \phi \left(n \right)$, and $e$ and$\phi \left(n \right)$ are mutually prime.

5) Compute the private key index $d$ satisfying $d \ast e \equiv  1 \left( Mod\ \phi \left(n \right) \right)$. $d$ is the multiplicative inverse of $e$ to $\phi \left(n \right)$.

After the key computation is complete, $n$ and $e$ are disclosed as the public key, where $n$ is the modulus and $e$ is the public key index. Convert the plaintext message $M$ to an integer $m$ with $0 < m < n$. Calculate the ciphertext$ C=m^{e} \left(Mod\ n \right)$. $C$ is the encrypted message. After receiving the ciphertext $C$, decrypt it using private key exponent $d$. Compute the plaintext message $M = C^{d} \left(Mod\ n\right)$. $m$ is the original plaintext message.

\subsection{Parallel Acceleration}
After deploying the large model separately from the client and the server, the server node will bear most of the training cost, and according to the difference in the computing power of the server node, the FL-GLM framework supports two training strategies: serial training and parallel training.
If the server node has limited computing resources and can hardly afford a large batch size, serial training is a more suitable choice.
As shown in Figure \ref{fig:compare}(b), during serial training, the server interacts with only one of the clients, and when one client completes the training, the training process for the next client is started. After completing the training, the parameters of multiple client models need to be averaged. Serial training is time-consuming, but one-to-one communication requires less communication, thread processing, and server processing power and is suitable for training scenarios with limited server capacity.

\begin{figure}[!t]
    \centering
    \includegraphics[width=1.0\linewidth]{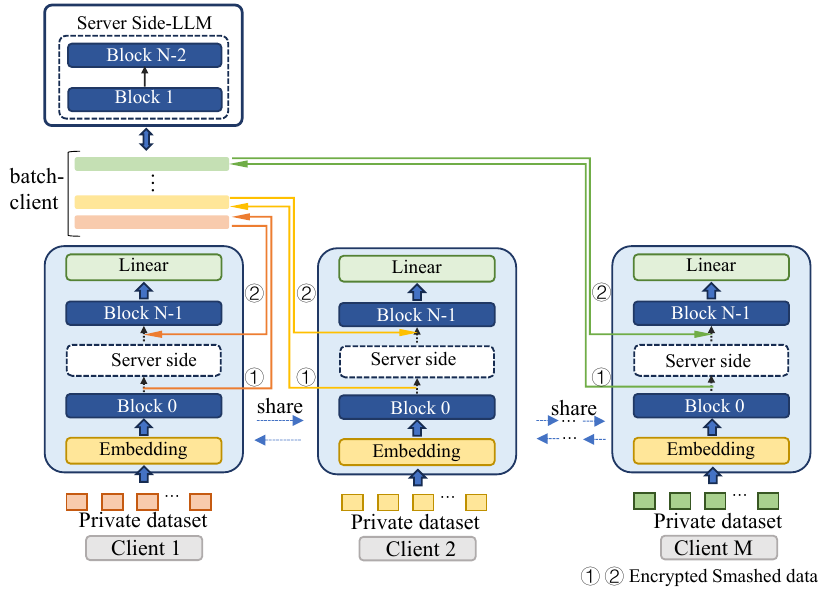}
    \caption{FL-GLM with client-batch parallel training. 
    }
    \label{fig:client-batch parallel framework}
\end{figure}

\begin{figure}[!t]
    \centering
    \includegraphics[width=1.0\linewidth]{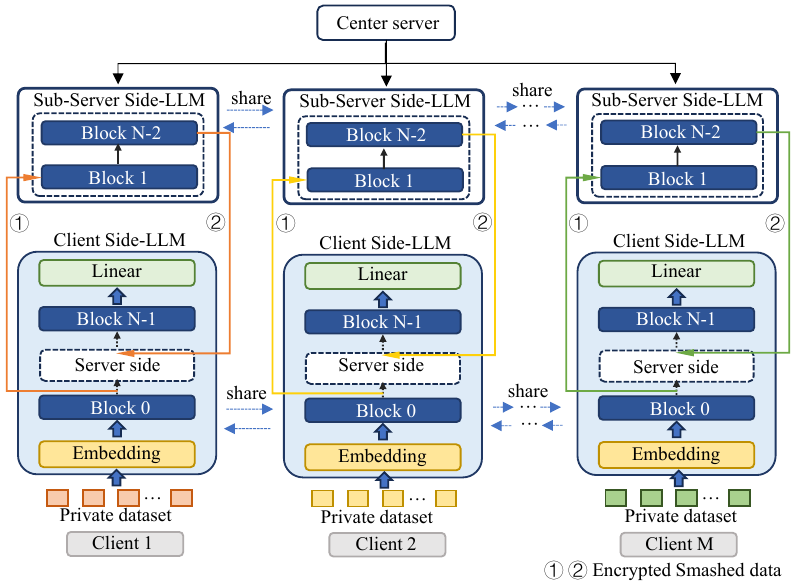}
    \caption{FL-GLM with server-hierarchical parallel. 
    }
    \label{fig:server-hierarchical framework}
\end{figure}

Since the special structure of split learning does not allow smashed data from multiple clients to be averaged, which will result in features and labels not being aligned and a substantial decrease in model performance, two parallel training strategies are designed in the FL-GLM framework.
As shown in Figure~\ref{fig:client-batch parallel framework}, the first strategy is to stack the smashed data from different clients during parallel training as a set of data to expand the batch for collaborative training.
Take clients' batch size=1 as an example; the number of clients is $M$, and in each round of training, every client sends smashed data of size {seqlength, batchsize=1, hiddensize} to the server, and the data received by the server will be integrated into a tensor with batch size $M$ for subsequent training. The second parallel strategy is shown in Figure~\ref{fig:server-hierarchical framework}.
Each client model will correspond to a server-side model, and the server node will run multiple models simultaneously, which can alleviate the threading problem in one-to-many communication to a certain extent.
The server-side model parameters and client-side parameters are averaged at the end of the training period. 

\section{Experiments}

In order to demonstrate the performance of chatGLM model within the federated learning framework(FL-GLM), we conduct experiments using the same benchmarks as those used in GLM model~\citep{glm}.

\begin{table*}[!t]
    \begin{center}
        \small
        \begin{adjustbox}{width=\textwidth}
        \begin{tabular}{lcccccccccc}
            \toprule
             \multirow {2}*{Model}& \multirow {2}*{Model Size}& \textbf{ReCoRD} & \textbf{COPA} & \textbf{WSC} & \textbf{RTE} & \textbf{BoolQ} & \textbf{WiC} & \textbf{CB} & \textbf{MultiRC} &\multirow{2}*{\textbf{Avg}} \\
             &  & F1/Acc. & Acc. &Acc. & Acc. & Acc. & Acc. & F1/Acc. &F1a/EM & \\  \hline
            $\text{T5}_{\text{large}}$~\citep{glm} & 770M & 85.7/85.0&78.0&84.6&84.8&84.3&71.6&96.4/98.2&80.9/46.6&81.2\\
            
            $\text{BART}_{\text{Large}}$~\citep{glm} & 409M & 88.3/87.8&60.0&65.4&84.5&84.3&69.0&90.5/92.9&81.8/48.0&76.0 \\
            
            $\text{RoBERTa}_{\text{Large}}$~\citep{glm} & 335M & 89.0/88.4&90.0&63.5&87.0&86.1&72.6&96.1/94.6&84.4/52.9&81.5 \\
            $\text{GLM}_{\text{RoBERTa}}$~\citep{glm} & 335M & 89.6/89.0&82.0&83.7&87.7&84.7&71.2&98.7/98.2&82.4/50.1&82.9\\
            ChatGLM-6B~\citep{zeng2022glm} & 6B & 80.2/78.7  &85.0  & 71.2 & 81.6 & 83.4 & 71.0 & 85.7/83.9 & 78.2/45.6 & 79.6 \\
            \hdashline
            FL-GLM & 6B & 79.8/78.4 &85.0 &71.2 &80.1&81.9&69.6&85.7/83.9&79.3/46.1&79.1  \\
            \bottomrule
            
        \end{tabular}
        \end{adjustbox}
    \end{center}
    \caption{\label{tb:resultSuper} Results on the SuperGLUE dev set.}
\end{table*}

\begin{table*}[!t]
     \small
    \begin{center}
    \begin{adjustbox}{width=\textwidth}
        \begin{tabular}{lcccccccc}
            \toprule
             \multirow {2}*{\textbf{Model}}& \multirow {2}*{Model Size}& \multicolumn{3}{c}{\textbf{CNN/DailyMail}} &\multicolumn{3}{c}{\textbf{XSum}} \\
             &  & ROUGE-1 & ROUGE-2 & ROUGE-L & ROUGE-1 & ROUGE-2 & ROUGE-L \\  \hline
            BERTSumAbs~\citep{liu2019text} & 110M & 41.7 &19.4 &38.8& 38.8& 16.3& 31.2\\
            $\text{UniLMv2}_{\text{Base}}$~\citep{bao2020unilmv2}& 110M & 43.2& 20.4& 40.1& 44.0& 21.1& 36.1 \\
            $\text{T5}_{\text{Large}}$~\citep{raffel2020exploring}&  770M& 42.5 &20.7 &39.8 &40.9 &17.3 &33.0 \\
            $\text{BART}_{\text{Large}} $ ~\citep{lewis2020bart}& 409M & 44.2& 21.3& 40.9& 45.1 &22.3 &37.3 \\
           $\text{GLM}_{\text{RoBERTa}} $~\citep{glm} & 335M & 43.8 &21.0 & 40.5 & 45.5 & 23.5 & 37.3 \\
           ChatGLM-6B~\citep{zeng2022glm}& 6B &  40.4 &17.0&28.0 &37.6& 12.5 &30.1 \\
            \hdashline
            FL-GLM & 6B & 39.6 & 16.9 & 28.0 & 37.0 & 11.9 & 29.4 \\
            \bottomrule
            
        \end{tabular}
        \end{adjustbox}
    \end{center}
    \caption{\label{tb:resultCNN} Results of abstractive summarization on the CNN/DailyMail and XSum test sets.}
\end{table*}

\subsection{Experimental Settings}
We first introduce some empirical settings, including datasets, evaluation metrics, baselines and parameter settings for FL-GLM.

\subsubsection{Dataset} 
For a fair comparison with centralized chatGLM-6B, we test our model on the SuperGLUE~\citep{wang2019superglue} benchmark for NLU tasks, and on CNN/DailyMail and XSum datasets for abstractive summarization tasks.

The SuperGLUE benchmark is a collection of challenging NLU tasks designed to evaluate the performance and capabilities of state-of-the-art language models. It consists of eight diverse tasks, i.e., ReCoRD, COPA, WSC, RTE, BoolQ, WiC, CB, and MultiRC, each representing a different aspect of language understanding. The details of the SuperGLUE benchmark can be seen in Appendix~\ref{sec:datasets detail}.
Following GLM~\citep{glm}, we formulate these tasks as blank infilling tasks. Specifically, given a labeled example $(x,y)$, we rewrite the input $x$ as a closed question $q(x)$ through a mask token [M] and rewrite output $y$ as an answer $a(y)$.

For abstractive summarization tasks, we append a mask token [M] at the end of the given context as input and treat the summary as output. Then the model generates the summary autoregressively.

\subsubsection{Metrics} 
Since the NLU tasks are reformulated as blank infilling tasks, the model performance can be evaluated using the generated probability of the ground-truth answer $a(y)$. For the RTE, BoolQ, WiC, CB, and MultiRC datasets, the generated answer may contain a single word. Therefore, we compute the logit of the corresponding answer token as the evaluation score, defined as:

\begin{equation*}
    p(y|x)=\frac{p(a(y)|q(x)}{ \sum_{y^{'} \in Y} p(a(y^{'})|q(x))},
\end{equation*}
where $Y$ is the ground-truth label set.

For the ReCoRD, COPA, and WSC datasets, the answers may contain multiple words; therefore, we compute the sum of the log-probabilities of the answer tokens as the evaluation metrics, which is defined as

\begin{equation*}
    s(y|x)=\sum_{t=1}^{|L_{y}| }\text{log} \text{P}(y_{t }|y_{1}...y_{t-1},x;\theta ).
\end{equation*}

For the summarization task, we use ROUGE-1, ROUGE-2, and ROUGE-L as quantitative metrics, which are widely used in NLP tasks~\citep{liu2021topic,chen2020multi,fang2022spoken}. 

\begin{table*}[!t]
    \begin{center}
    \begin{adjustbox}{width=\textwidth}
        \begin{tabular}{lcccccccccr}
        \toprule
         \textbf{Strategy} & \textbf{Centralized} &  \textbf{serial} & \multicolumn{3}{c}{\textbf{client-batch parallel}} & \multicolumn{4}{c}{\textbf{server-hierarchical}} &  \\ \hline
         num. of clients & None & 2 & 2     & 4     & 8     & 2     & 3     & 5     & 10    &  \\
         \hdashline
        time(s) & 166.4±9.2 & 175.2±10.1 & 85.3±4.1 & 43.0±2.5 & 22.5±1.7 & 87.3±4.9 & 65.5±3.2 & 34.5±1.9 & 17.3±0.9 &  \\
       \bottomrule     
       \end{tabular}%
       \end{adjustbox}
    \end{center}
    \caption{\label{tb:communication} Comparison of training time between different training strategies}   
\end{table*}%

\subsubsection{Baselines} 

We apply FL-GLM to ChatGLM-6B model\footnote{\href{https://github.com/THUDM/ChatGLM-6B}{https://github.com/THUDM/ChatGLM-6B}}, who is an open-source pre-trained language model with 6 billion parameters and building upon the General Language Model(GLM-130B)~\citep{zeng2022glm,glm}. 
Notely, our framework is not limited to ChatGLM but can be widely applied to different LLMs (such as Llama2). We use ChatGLM as a representative model to demonstrate that our framework does not significantly degrade model performance.  Considering that our future applications will mainly focus on the Chinese domain, we chose ChatGLM-6B, which has been extensively aligned with human in the Chinese domain. Additionally, the ChatGLM-6B model offers a breakthrough scaling property that enables efficient inference on a single RTX 3060 (12GB) GPU through INT4 quantization. This property is especially valuable in resource-constrained scenarios, allowing for cost-effective computation on affordable GPUs.

For a fair comparison with ChatGLM-6B, following GLM, we use 7 baselines, including $\text{T5}_{\text{large}}$~\citep{raffel2020exploring}, $\text{BART}_{\text{Large}}$~\citep{lewis2020bart}, $\text{RoBERTa}_{\text{Large}}$~\citep{liu2019roberta}, $\text{GLM}_{\text{RoBERTa}}$~\citep{glm}, BERTSumAbs~\citep{liu2019text}, $\text{UniLMv2}_{\text{Base}}$~\citep{bao2020unilmv2} and ChatGLM-6B~\citep{zeng2022glm}.

\subsubsection{Parameter Settings}
We utilize the open-source ChatGLM-6B model as the basement model for the FL-GLM model. It has 28-layer transformer blocks, 4096 hidden-size, and 32 self-attention heads. We utilize P-tuning v2 for more efficient fine-tuning on downstream tasks. Experiments are conducted on 2, 3, 5, and 10 clients with NVIDIA A100 GPUs, 40GB RAM per client, and one server with one NVIDIA A100 GPU and 40GB RAM. We generate RSA public and private keys at the beginning of FL and then pass the public keys between server and client. During the FL process, the keys remain unchanged, and after a certain number (hyper-parameter) of rounds of training, we regenerate and share the keys. Our experiments are conducted with communication simulated on the same host, but not in a for-loop manner; rather, we coordinated information with Flower tool~\footnote{https://github.com/mher/flower}.
In order to make a fair comparison between our FL-GLM model and ChatGLM-6B, we used a batch size of one, a learning rate of 2e-2 with the Adam optimizer, and adjusted the number of training epochs and maximum sequence length according to different datasets without using warmup or weight decay. The code will be released when this paper is accepted.

\subsection{Experimental Results}
In this section, we demonstrate our experiment results on SuperGLUE benchmark, CNN/DialyMail and XSum datasets.
\subsubsection{Metric-based Evaluation}
The quantitative evaluation results on SuperGLUE are shown in Table~\ref{tb:resultSuper}. 
 
From the results, we can see that the recent large language models, such as ChatGLM-6B outperform the traditional pre-training models, showing the effectiveness of human-aligned language models for NLU tasks. As a distributed learning pattern, our FL-GLM model performs a little worse than the basement model, ChatGLB-6B. Take the accuracy of the ReCoRD, RTE, BoolQ, and Wic datasets. For example, our FL-GLM model obtains 78.4, 81.6, 81.9, and 69.6, respectively, which is lower than the centralized ChatGLB-6B model in the acceptable range, i.e., 0.3, 1.5, 1.5, and 1.4.  

From the results on CNN/DialyMail and XSum datasets in Table~\ref{tb:resultCNN}, shiFL-GLM can obtain 39.6 ROUGE-1, 16.9 ROUGE-2, and 28.0 ROUGE-L on the CNN/DailyMail dataset, 37.0 ROUGE-1, 11.9 ROUGE-2, and 29.4 ROUGE-L on the XSum dataset. Not more than 1.0 lower than the results of the centralized ChatGLM-6B model.

In conclusion, our FL-GLM model has comparable ability to understand language and generate relevant summary with centralized models.

\begin{figure*}[!t]
    \centering
    \includegraphics[width=1.0\linewidth]{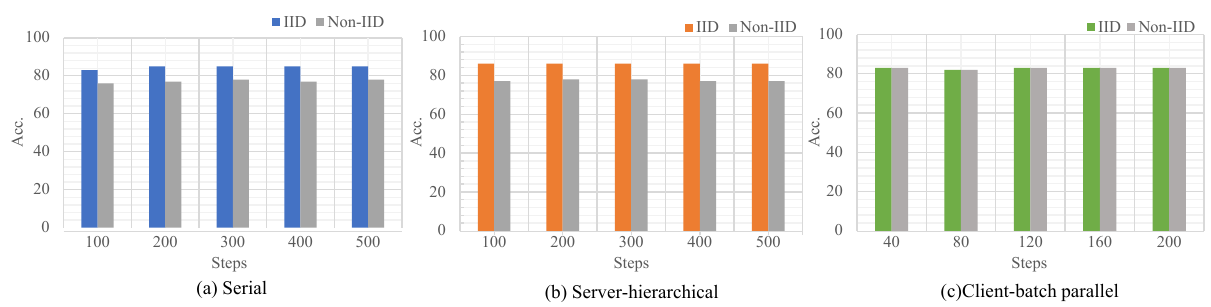}
    \caption{{Impact of IID and non-IID of COPA dataset on  FL-GLM.}}
    \label{fig:iid result}
\end{figure*}

\subsection{Analysis}

An analysis is conducted including training efficiency, impact of Data non-IID and prove the security of FL-GLM. We also conducted the experiments to analysis the impact of average period (Appendix~\ref{app:average}) and the impact of participants(Appendix~\ref{app:part}).

\subsubsection{Training Efficiency}
To further investigate the impact of our speedup optimization mechanism on the training cost, we tested the average training duration of the FL-GLM model under three training strategies: serial, client-batch, and server-hierarchical. We randomly selected 1000 data points from the ReCoRD dataset for communication cost analysis experiments. We tested 10 times and took the mean and standard deviation of the total communication time, as shown in Table ~\ref{tb:communication}.From the results, we can see that the time consumed in serial training mode with 1000 data points is close to that of centralized training, while parallel training can significantly improve the training time, which is directly proportional to the number of clients. 

Furthermore, we measured training communication time between two computers (server and client) within the same LAN with a bandwidth of 1100MB$/$s. Results indicate that training across two machines in the LAN is approximately five times slower than centralized training (centralized training: 0.91s$/$step, stimulate training FL-GLM in single machine with two GPUs: 1.79s$/$step, training FL-GLM with two machines: 4.83s$/$step).

\subsubsection{{Impact of IID and non-IID}}

Whether the data satisfy the independent and identically distributed (IID) assumption is one of the important challenges of federated learning. To test the effect of data distribution on the performance of FL-GLM, we conducted the following experiments. We selected the COPA dataset from the SuperGLUE benchmark, which is a binary classification dataset for textual causal judgment. It contains 400 training samples, with 195 labeled as 0 and 205 labeled as 1.

To simplify the analysis, we assumed the existence of two clients. After sampling with the independent and identically distributed (IID) method, the dataset was divided into sub-datasets A and B. Sub-dataset A contains 97 samples labeled as 0 and 102 samples labeled as 1, while sub-dataset B contains the remaining samples. Then, we applied a non-IID sampling method to divide the datasets into sub-datasets A' and B'. Sub-dataset A' contains 195 samples labeled as 0 and 5 samples labeled as 1, while B' contains 200 samples labeled as 1.

The experimental results under three training strategies are shown in Figure~\ref{fig:iid result}. In the case of non-IID data, due to the issue of data heterogeneity, the model performance of fine-tuning training using both serial training strategy~\ref{fig:iid result}(a) and server-parallel strategy~\ref{fig:iid result}(b) decreases by approximately 7\%. However, client-batch parallel training~\ref{fig:iid result}(c) is not significantly affected by the data distribution. This is because during client-batch parallel training, the data features from each client are stacked into batches and sent to the server, allowing most of model parameters on the server side to sufficiently learn the data features, to some extent mitigates the performance loss caused by non-IID data.

\subsubsection{{Security Analysis}}
Theoretical proof of the security of split learning is challenging. ~\citep{pasquini2021unleashing} propose an inference attack method FSHA for feature data security in split learning, where a malicious server restores the training dataset by hijacking the client's output data, which is validated in the field of image recognition, and is able to restore the client's training dataset effectively. Inspired by this method, we conduct security analysis experiments to indirectly demonstrate the security of FL-GLM.

An important prerequisite for FSHA is that the malicious server has a shadow dataset with the same domain and task as the dataset held by the attacked party. However, in the private data domain, the data are all held by the training participants and protected by legal regulations, and the server side in the FL-GLM framework cannot obtain the same domain data under normal circumstances. sSo we consider the extreme case where, in serial training mode, at least one client colludes with the server to share its private data, ${D_{priv}}^1$, with the server for the purpose of training an attack model. Let $F$ be the first part of the model held by the malicious client. The malicious server-side constructs the model $F^{-1}$ for attacking and utilizes ${D_{priv}}^1$ to train $F^{-1}$.  During the attacking phase, the malicious server hijacks the smashed data outputted by the attacked client, denoted as $f^{~}$, and utilizes $F^{-1}$ to inference the privacy data ${D_{priv}}^2$ held by the attacked client.The method is validated on the BoolQ dataset.

The experimental results are shown in Table \ref{tb:security}. When the client only has the embedding layer like FedBert model, $F^{-1}$ is a single Linear layer, the attack model can achieve a BLEU-4 score of 28.570 and a ROUGE-1 score of 33.290, while in the FL-GLM framework, where the client contains the embedding layer and an LLM-Block, $F^{-1}$ is a single layer Transfomer, all the metrics of the attack model are all close to 0. Therefore, the security of FL-GLM could be proven in experiments. Additionally, we find that the attack metrics' performance of a single-block Transformer is similar to that of a multi-block Transformer. Therefore, the optimal split point, based on experimental results, might be a single-block Transformer, even though it is challenging to prove theoretically.

\begin{table}[t]
  \begin{center}
    \begin{adjustbox}{width=0.48\textwidth}
        \begin{tabular}{cccccc}
        \toprule
    \textbf{$F$} & \textbf{$F^{-1}$}& \textbf{Rouge-1} & \textbf{Rouge-2} & \textbf{Rouge-l} & \textbf{Bleu-4} \\
    \midrule
    Embedding(FebBert) & Linear & 33.29 & 7.053 & 26.732 & 28.57 \\
    \midrule
    FL-GLM client-side part A & Transformer & 0.135 & 0.002 & 0.473 & 0.335 \\
    \bottomrule
    \end{tabular}%
       \end{adjustbox}
    \end{center}
    \caption{\label{tb:security} Security Analysis }   
\end{table}%

\subsubsection{More Basement Model}

To further validate the practicality and generalizability of our FL-GLM framework, we incorporated Llama2-7B-Chat~\citep{touvron2023llama} as a baseline model. We split Llama2-7B-Chat according to the FL-GLM framework, resulting in a variant denoted as FL-Llama. 
The performance evaluation experiment was conducted utilizing six datasets from the SuperGLUE benchmark with serial training strategy and utilizing the LoRA to fine-tune. 
Note that Llama2-7B-Chat supports instruction fine-tuning, leading us to adapt the input prompts for each dataset in accordance with the Meta's official recommended instruction templates, rather than employing the Cloze question templates in Appendix~\ref{sec:datasets detail}.
The experimental results are presented in Table~\ref{tb:test llama}, which indicate that our proposed framework could achieve comparable accuracy metrics, and suggest that the FL-GLM framework is agnostic to the base model type and does not significantly affect performance.



Furthermore, we validated the FL-GLM framework on the Chinese medical dataset Huatuo-26M~\citep{li2023huatuo26m}, referring to the Llama2-Chinese-7B-Chat-based model as FL-Chinese-Llama.
The full Huatuo-26M dataset contains 2,623,904 Q\&A pairs in the training dataset and 264,041 Q\&A pairs in the test dataset. We randomly sampled 3,000 Q\&A pairs from the full dataset for training and 300 Q\&A pairs from the test set for evaluation. 
It should be noted that, since the limited Chinese language capability of Llama2-7B-chat, we chose Llama2-Chinese-7B-Chat~\footnote{https://huggingface.co/FlagAlpha/Llama2-Chinese-7B-Chat} as the baseline model for this experiment, which has been fine-tuned on a large Chinese corpus. 
As illustrated in Table~\ref{tb:test huatuo}, the proposed training framework maintains model performance that is closely comparable to centralized training. Specifically, the performance metrics of FL-GLM exhibit a degradation of no more than 0.72 when compared to ChatGLM-6B, while FL-Chinese-Llama shows a decline of no more than 0.38 when compared to Llama2-Chinese-7B-Chat. Whereas, the inter-model comparisons show that the Chinese conversational capabilities of ChatGLM significantly surpass those of Llama2-Chinese-7B-Chat. Therefore, we recommend using ChatGLM as the base model for FL-GLM framework in Chinese application scenarios.

\begin{table}[!t]
  \centering  
    \begin{adjustbox}{width=0.48\textwidth}
    \begin{tabular}{ccccccc}
    \toprule
    Model & CB & COPA & RTE & WiC &BoolQ & WSC \\
    \midrule
    Llama2-7B-Chat & 85.71 & 75.00 & 73.29 & 70.85 & 83.79 & 68.27 \\
    FL-Llama(ours) & 85.71 & 75.00 & 73.29 & 70.83 & 82.35 & 68.37 \\
    \bottomrule
    \end{tabular}%
    \end{adjustbox}
    \caption{\label{tb:test llama} FL-GLM's performance on Llama2-7B-Chat}
  
\end{table}%

\begin{table}[!t]
  \centering  
    \begin{adjustbox}{width=0.48\textwidth}
    \begin{tabular}{ccccc}
    \toprule
    Model & BLEU-4 & ROUGE-1 & ROUGE-2 &ROUGE-L \\
    \midrule
    ChatGLM-6B & 9.32  & 29.94 & 8.89  & 24.07 \\
    FL-GLM(OURS) & 8.95  & 29.22 & 8.61   & 23.40 \\
    Llama2-Chinese-7B-Chat & 3.07     & 19.42     & 2.95     & 15.08 \\
    FL-Chinese-Llama(ours)  & 3.04     & 19.04     & 2.79     & 14.85 \\
     \bottomrule
    \end{tabular}%
    \end{adjustbox}
    \caption{\label{tb:test huatuo} Results on sampled Huatuo-26M test set.}
\end{table}%

\section{Conclusions}

To address the challenge of distributed training of LLMs with limited client computational resources, we propose to utilize the split learning method to segment the generative model. We place the input and output blocks locally on client devices, while the remaining primary model parameters are centralized on a server with ample computational resources. We secure client-server information transfers with encryption methods. To enhance training efficiency, we suggest selecting the client-batch and server-hierarchical acceleration optimization methods based on the server's actual computational capacity, thereby enabling parallel training. This distributed architecture not only ensures that user private data remains on local devices but also effectively reduces the training time, making it more suitable for the scale and complexity of LLMs. In the future, we contemplate employing more advanced privacy-preserving techniques, such as differential privacy, to safeguard the data transmitted from clients, enabling the application of large language models in privacy-sensitive scenarios.

\section*{Limitations}

FL-GLM was evaluated on the SuperGLUE benchmark, CNN/DailyMail and XSum datasets, and despite achieving results close to those of the centralized tests, it is still constrained by the privacy-utility trade-off, and we would like to further optimize the communication consumption of the current distributed training framework and achieve even better model efficacy. In addition, our framework is currently limited to ChatGLM-6B. Future work will extend FL-GLM to different LLMs, such as Llama, to demonstrate its adaptability and wider applicability.

\section*{Ethical Considerations}
We propose a federated learning framework named FL-GLM, which aims to use private data to train LLM with considerations of prevent data privacy leakage. Our data originates from open-source NLU and NLG projects, adhering to their license limitations and public benchmarks. Moreover, we emulate a distributed data storage environment using open-source datasets, ensuring the exclusion of private data. We affirm our societal contribution without causing harm.




\begin{thebibliography}{36}
\expandafter\ifx\csname natexlab\endcsname\relax\def\natexlab#1{#1}\fi

\bibitem[{Abedi and Khan(2020)}]{abedi2020fedsl}
Ali Abedi and Shehroz~S Khan. 2020.
\newblock Fedsl: Federated split learning on distributed sequential data in recurrent neural networks.
\newblock \emph{arXiv preprint arXiv:2011.03180}.

\bibitem[{Abuadbba et~al.(2020)Abuadbba, Kim, Kim, Thapa, Camtepe, Gao, Kim, and Nepal}]{abuadbba2020can}
Sharif Abuadbba, Kyuyeon Kim, Minki Kim, Chandra Thapa, Seyit~A Camtepe, Yansong Gao, Hyoungshick Kim, and Surya Nepal. 2020.
\newblock Can we use split learning on 1d cnn models for privacy preserving training?
\newblock In \emph{Proceedings of the 15th ACM Asia Conference on Computer and Communications Security}, pages 305--318.

\bibitem[{Asnani et~al.(2023)Asnani, Yin, Hassner, and Liu}]{asnani2023reverse}
Vishal Asnani, Xi~Yin, Tal Hassner, and Xiaoming Liu. 2023.
\newblock Reverse engineering of generative models: Inferring model hyperparameters from generated images.
\newblock \emph{IEEE Transactions on Pattern Analysis and Machine Intelligence}.

\bibitem[{Bao et~al.(2020)Bao, Dong, Wei, Wang, Yang, Liu, Wang, Piao, Gao, Zhou et~al.}]{bao2020unilmv2}
Hangbo Bao, Li~Dong, Furu Wei, Wenhui Wang, Nan Yang, Xiaodong Liu, Yu~Wang, Songhao Piao, Jianfeng Gao, Ming Zhou, et~al. 2020.
\newblock Unilmv2: pseudo-masked language models for unified language model pre-training.
\newblock In \emph{Proceedings of the 37th International Conference on Machine Learning}, pages 642--652.

\bibitem[{Chen and Yang(2020)}]{chen2020multi}
Jiaao Chen and Diyi Yang. 2020.
\newblock Multi-view sequence-to-sequence models with conversational structure for abstractive dialogue summarization.
\newblock In \emph{Proceedings of the 2020 Conference on Empirical Methods in Natural Language Processing (EMNLP)}, pages 4106--4118.

\bibitem[{Chen et~al.(2019)Chen, Mathews, Ouyang, and Beaufays}]{chen2019federated}
Mingqing Chen, Rajiv Mathews, Tom Ouyang, and Fran{\c{c}}oise Beaufays. 2019.
\newblock Federated learning of out-of-vocabulary words.
\newblock \emph{arXiv preprint arXiv:1903.10635}.

\bibitem[{Du et~al.(2022)Du, Qian, Liu, Ding, Qiu, Yang, and Tang}]{glm}
Zhengxiao Du, Yujie Qian, Xiao Liu, Ming Ding, Jiezhong Qiu, Zhilin Yang, and Jie Tang. 2022.
\newblock Glm: General language model pretraining with autoregressive blank infilling.
\newblock In \emph{Proceedings of the 60th Annual Meeting of the Association for Computational Linguistics (Volume 1: Long Papers)}, pages 320--335.

\bibitem[{Fang et~al.(2022)Fang, Zhang, Chen, Ding, Long, Lan, and Zhou}]{fang2022spoken}
Yue Fang, Hainan Zhang, Hongshen Chen, Zhuoye Ding, Bo~Long, Yanyan Lan, and Yanquan Zhou. 2022.
\newblock From spoken dialogue to formal summary: An utterance rewriting for dialogue summarization.
\newblock In \emph{Proceedings of the 2022 Conference of the North American Chapter of the Association for Computational Linguistics: Human Language Technologies}, pages 3859--3869.

\bibitem[{Gupta et~al.(2022)Gupta, Huang, Zhong, Gao, Li, and Chen}]{gupta2022recovering}
Samyak Gupta, Yangsibo Huang, Zexuan Zhong, Tianyu Gao, Kai Li, and Danqi Chen. 2022.
\newblock Recovering private text in federated learning of language models.
\newblock \emph{Advances in Neural Information Processing Systems}, 35:8130--8143.

\bibitem[{Hard et~al.(2018)Hard, Rao, Mathews, Ramaswamy, Beaufays, Augenstein, Eichner, Kiddon, and Ramage}]{hard2018federated}
Andrew Hard, Kanishka Rao, Rajiv Mathews, Swaroop Ramaswamy, Fran{\c{c}}oise Beaufays, Sean Augenstein, Hubert Eichner, Chlo{\'e} Kiddon, and Daniel Ramage. 2018.
\newblock Federated learning for mobile keyboard prediction.
\newblock \emph{arXiv preprint arXiv:1811.03604}.

\bibitem[{Huang et~al.(2020)Huang, Song, Chen, Li, and Arora}]{huang2020texthide}
Yangsibo Huang, Zhao Song, Danqi Chen, Kai Li, and Sanjeev Arora. 2020.
\newblock Texthide: Tackling data privacy in language understanding tasks.
\newblock In \emph{Findings of the Association for Computational Linguistics: EMNLP 2020}, pages 1368--1382.

\bibitem[{Jalalirad et~al.(2019)Jalalirad, Scavuzzo, Capota, and Sprague}]{jalalirad2019simple}
Amir Jalalirad, Marco Scavuzzo, Catalin Capota, and Michael Sprague. 2019.
\newblock A simple and efficient federated recommender system.
\newblock In \emph{Proceedings of the 6th IEEE/ACM international conference on big data computing, applications and technologies}, pages 53--58.

\bibitem[{Ji et~al.(2019)Ji, Pan, Long, Li, Jiang, and Huang}]{ji2019learning}
Shaoxiong Ji, Shirui Pan, Guodong Long, Xue Li, Jing Jiang, and Zi~Huang. 2019.
\newblock Learning private neural language modeling with attentive aggregation.
\newblock In \emph{2019 International joint conference on neural networks (IJCNN)}, pages 1--8. IEEE.

\bibitem[{Lewis et~al.(2020)Lewis, Liu, Goyal, Ghazvininejad, Mohamed, Levy, Stoyanov, and Zettlemoyer}]{lewis2020bart}
Mike Lewis, Yinhan Liu, Naman Goyal, Marjan Ghazvininejad, Abdelrahman Mohamed, Omer Levy, Veselin Stoyanov, and Luke Zettlemoyer. 2020.
\newblock Bart: Denoising sequence-to-sequence pre-training for natural language generation, translation, and comprehension.
\newblock In \emph{Proceedings of the 58th Annual Meeting of the Association for Computational Linguistics}, pages 7871--7880.

\bibitem[{Li et~al.(2020{\natexlab{a}})Li, Fan, Tse, and Lin}]{li2020review}
Li~Li, Yuxi Fan, Mike Tse, and Kuo-Yi Lin. 2020{\natexlab{a}}.
\newblock A review of applications in federated learning.
\newblock \emph{Computers \& Industrial Engineering}, 149:106854.

\bibitem[{Li et~al.(2020{\natexlab{b}})Li, Sahu, Zaheer, Sanjabi, Talwalkar, and Smith}]{li2020federated}
Tian Li, Anit~Kumar Sahu, Manzil Zaheer, Maziar Sanjabi, Ameet Talwalkar, and Virginia Smith. 2020{\natexlab{b}}.
\newblock Federated optimization in heterogeneous networks.
\newblock \emph{Proceedings of Machine learning and systems}, 2:429--450.

\bibitem[{Li et~al.(2021)Li, Tramer, Liang, and Hashimoto}]{li2021large}
Xuechen Li, Florian Tramer, Percy Liang, and Tatsunori Hashimoto. 2021.
\newblock Large language models can be strong differentially private learners.
\newblock In \emph{International Conference on Learning Representations}.

\bibitem[{Liu et~al.(2021{\natexlab{a}})Liu, Zou, Zhang, Chen, Ding, Yuan, and Wang}]{liu2021topic}
Junpeng Liu, Yanyan Zou, Hainan Zhang, Hongshen Chen, Zhuoye Ding, Caixia Yuan, and Xiaojie Wang. 2021{\natexlab{a}}.
\newblock Topic-aware contrastive learning for abstractive dialogue summarization.
\newblock In \emph{Findings of the Association for Computational Linguistics: EMNLP 2021}, pages 1229--1243.

\bibitem[{Liu et~al.(2021{\natexlab{b}})Liu, Ji, Fu, Tam, Du, Yang, and Tang}]{liu2021p}
Xiao Liu, Kaixuan Ji, Yicheng Fu, Weng~Lam Tam, Zhengxiao Du, Zhilin Yang, and Jie Tang. 2021{\natexlab{b}}.
\newblock P-tuning v2: Prompt tuning can be comparable to fine-tuning universally across scales and tasks.
\newblock \emph{arXiv preprint arXiv:2110.07602}.

\bibitem[{Liu et~al.(2021{\natexlab{c}})Liu, Zheng, Du, Ding, Qian, Yang, and Tang}]{liu2021gpt}
Xiao Liu, Yanan Zheng, Zhengxiao Du, Ming Ding, Yujie Qian, Zhilin Yang, and Jie Tang. 2021{\natexlab{c}}.
\newblock Gpt understands, too.
\newblock \emph{arXiv e-prints}, pages arXiv--2103.

\bibitem[{Liu and Lapata(2019)}]{liu2019text}
Yang Liu and Mirella Lapata. 2019.
\newblock Text summarization with pretrained encoders.
\newblock In \emph{Proceedings of the 2019 Conference on Empirical Methods in Natural Language Processing and the 9th International Joint Conference on Natural Language Processing (EMNLP-IJCNLP)}, pages 3730--3740.

\bibitem[{Liu et~al.(2019)Liu, Ott, Goyal, Du, Joshi, Chen, Levy, Lewis, Zettlemoyer, and Stoyanov}]{liu2019roberta}
Yinhan Liu, Myle Ott, Naman Goyal, Jingfei Du, Mandar Joshi, Danqi Chen, Omer Levy, Mike Lewis, Luke Zettlemoyer, and Veselin Stoyanov. 2019.
\newblock Roberta: A robustly optimized bert pretraining approach.
\newblock \emph{arXiv preprint arXiv:1907.11692}.

\bibitem[{Matsubara and Levorato(2020)}]{matsubara2020neural}
Y~Matsubara and M~Levorato. 2020.
\newblock Neural compression and filtering for edge-assisted real-time object detection in challenged networks.
\newblock In \emph{IEEE International Conference on Pattern Recognition (IEEE ICPR)}.

\bibitem[{McMahan et~al.(2017)McMahan, Moore, Ramage, Hampson, and y~Arcas}]{mcmahan2017communication}
Brendan McMahan, Eider Moore, Daniel Ramage, Seth Hampson, and Blaise~Aguera y~Arcas. 2017.
\newblock Communication-efficient learning of deep networks from decentralized data.
\newblock In \emph{Artificial intelligence and statistics}, pages 1273--1282. PMLR.

\bibitem[{Pasquini et~al.(2021)Pasquini, Ateniese, and Bernaschi}]{pasquini2021unleashing}
Dario Pasquini, Giuseppe Ateniese, and Massimo Bernaschi. 2021.
\newblock Unleashing the tiger: Inference attacks on split learning.
\newblock In \emph{Proceedings of the 2021 ACM SIGSAC Conference on Computer and Communications Security}, pages 2113--2129.

\bibitem[{Raffel et~al.(2020)Raffel, Shazeer, Roberts, Lee, Narang, Matena, Zhou, Li, and Liu}]{raffel2020exploring}
Colin Raffel, Noam Shazeer, Adam Roberts, Katherine Lee, Sharan Narang, Michael Matena, Yanqi Zhou, Wei Li, and Peter~J Liu. 2020.
\newblock Exploring the limits of transfer learning with a unified text-to-text transformer.
\newblock \emph{The Journal of Machine Learning Research}, 21(1):5485--5551.

\bibitem[{Rahman et~al.(2020)Rahman, Tout, Talhi, and Mourad}]{rahman2020internet}
Sawsan~Abdul Rahman, Hanine Tout, Chamseddine Talhi, and Azzam Mourad. 2020.
\newblock Internet of things intrusion detection: Centralized, on-device, or federated learning?
\newblock \emph{IEEE Network}, 34(6):310--317.

\bibitem[{Stremmel and Singh(2021)}]{stremmel2021pretraining}
Joel Stremmel and Arjun Singh. 2021.
\newblock Pretraining federated text models for next word prediction.
\newblock In \emph{Advances in Information and Communication: Proceedings of the 2021 Future of Information and Communication Conference (FICC), Volume 2}, pages 477--488. Springer.

\bibitem[{Thakkar et~al.(2020)Thakkar, Ramaswamy, Mathews, and Beaufays}]{thakkar2020understanding}
Om~Thakkar, Swaroop Ramaswamy, Rajiv Mathews, and Fran{\c{c}}oise Beaufays. 2020.
\newblock Understanding unintended memorization in federated learning.
\newblock \emph{arXiv preprint arXiv:2006.07490}.

\bibitem[{Thapa et~al.(2022)Thapa, Arachchige, Camtepe, and Sun}]{thapa2022splitfed}
Chandra Thapa, Pathum Chamikara~Mahawaga Arachchige, Seyit Camtepe, and Lichao Sun. 2022.
\newblock Splitfed: When federated learning meets split learning.
\newblock In \emph{Proceedings of the AAAI Conference on Artificial Intelligence}, volume~36, pages 8485--8493.

\bibitem[{Tian et~al.(2022)Tian, Wan, Lyu, Yao, Jin, and Sun}]{tian2022fedbert}
Yuanyishu Tian, Yao Wan, Lingjuan Lyu, Dezhong Yao, Hai Jin, and Lichao Sun. 2022.
\newblock Fedbert: When federated learning meets pre-training.
\newblock \emph{ACM Transactions on Intelligent Systems and Technology (TIST)}, 13(4):1--26.

\bibitem[{Wang et~al.(2019)Wang, Pruksachatkun, Nangia, Singh, Michael, Hill, Levy, and Bowman}]{wang2019superglue}
Alex Wang, Yada Pruksachatkun, Nikita Nangia, Amanpreet Singh, Julian Michael, Felix Hill, Omer Levy, and Samuel~R Bowman. 2019.
\newblock Superglue: a stickier benchmark for general-purpose language understanding systems.
\newblock In \emph{Proceedings of the 33rd International Conference on Neural Information Processing Systems}, pages 3266--3280.

\bibitem[{Touvron et~al.(2023)Touvron, Martin, Stone, Albert, Almahairi, Babaei, Bashlykov, Batra, Bhargava, Bhosale et~al.}]{touvron2023llama}
Hugo Touvron, Louis Martin, Kevin Stone, Peter Albert, Amjad Almahairi, Yasmine Babaei, Nikolay Bashlykov, Soumya Batra, Prajjwal Bhargava, Shruti Bhosale, et~al. 2023.
\newblock Llama 2: Open foundation and fine-tuned chat models.
\newblock \emph{arXiv preprint arXiv:2307.09288}.

\bibitem[{Yaldiz et~al.(2023)Yaldiz, Zhang, and Avestimehr}]{yaldiz2023secure}
Duygu~Nur Yaldiz, Tuo Zhang, and Salman Avestimehr. 2023.
\newblock Secure federated learning against model poisoning attacks via client filtering.
\newblock In \emph{ICLR 2023 Workshop on Backdoor Attacks and Defenses in Machine Learning}.

\bibitem[{Yu et~al.(2021)Yu, Naik, Backurs, Gopi, Inan, Kamath, Kulkarni, Lee, Manoel, Wutschitz et~al.}]{yu2021differentially}
Da~Yu, Saurabh Naik, Arturs Backurs, Sivakanth Gopi, Huseyin~A Inan, Gautam Kamath, Janardhan Kulkarni, Yin~Tat Lee, Andre Manoel, Lukas Wutschitz, et~al. 2021.
\newblock Differentially private fine-tuning of language models.
\newblock In \emph{International Conference on Learning Representations}.

\bibitem[{Li et~al.(2023)Li, Wang, Wu, Zhang, Xu, Fu, Tiwari, Wan, and Wang}]{li2023huatuo26m}
Jianquan Li, Xidong Wang, Xiangbo Wu, Zhiyi Zhang, Xiaolong Xu, Jie Fu, Prayag Tiwari, Xiang Wan, and Benyou Wang. 2023.
\newblock \href {http://arxiv.org/abs/2305.01526} {Huatuo-26m, a large-scale chinese medical qa dataset}.

\bibitem[{Zeng et~al.(2022)Zeng, Liu, Du, Wang, Lai, Ding, Yang, Xu, Zheng, Xia et~al.}]{zeng2022glm}
Aohan Zeng, Xiao Liu, Zhengxiao Du, Zihan Wang, Hanyu Lai, Ming Ding, Zhuoyi Yang, Yifan Xu, Wendi Zheng, Xiao Xia, et~al. 2022.
\newblock Glm-130b: An open bilingual pre-trained model.
\newblock In \emph{The Eleventh International Conference on Learning Representations}.

\bibitem[{Zhu et~al.(2019)Zhu, Liu, and Han}]{zhu2019deep}
Ligeng Zhu, Zhijian Liu, and Song Han. 2019.
\newblock Deep leakage from gradients.
\newblock In \emph{Proceedings of the 33rd International Conference on Neural Information Processing Systems}, pages 14774--14784.

\end{thebibliography}
\bibliographystyle{acl_natbib}

\appendix

\begin{table*}[!t]
\newcommand{\tabincell}[2]{\begin{tabular}{@{}#1@{}}#2\end{tabular}}
    \begin{center}
        \begin{adjustbox}{width=\textwidth}
        \begin{tabular}{lccc}
            \toprule
            Dataset & Task & Cloze Question & Answers \\ \hline
            ReCoRD & Question answering & [passage p] [cloze question q] & Answer candidates\\
            COPA& Causal reasoning & ``[choice c1]'' or``[choice c2]''? [premise p], so [M]. & c1/c2\\
            WSC & Coreference resolution& [sentence s] The pronoun `*p*' refers to [M]. & Noun n\\
            RTE & Textual entailment & ``[hypothesis h]''? [M]    ``[premise p]'' & \tabincell{l}{``yes''/``no''}\\
            BoolQ & Question answering & [passage p]. Question: q? Answer:[M]. & ``yes'' / ``no''\\
            WiC & Word sense disambiguation & \tabincell{l}{``[sentence s1]''/``[sentence s2]''Similar sense \\of [word w]? [M]. }&``yes''/``no'' \\
            CB & Textual entailment& ``[hypothesis h]''? [M], ``[premise p]'' &  \tabincell{l}{``yes''/``no''/``maybe''}\\
            MultiRC & Question answering &[passage p]. Question: q? Is it [answer a]? [M]. & ``yes''/``no'' \\            \bottomrule
            
        \end{tabular}
        \end{adjustbox}
    \end{center}
    \caption{\label{tb:superglueTask} Cloze questions and answers for the 8 SuperGLUE tasks}
\end{table*}

\section{P-tuning v2}\label{sec:model-details}

P-tuning v2 is proposed based on the p-tuning~\citep{liu2021gpt} algorithm, and its basic principle is to add a prompt of length $L_p$ as a learnable embedding, denoted as a prefix, to each LLM-Block's attention operation. Fine-tuning is done by freezing the model parameters and training only the prefix. In each LLM-Block, the corresponding prefix contains two parts: $ {prefix\_key} \in R^ {L\times B\times {N_h}\times {d_h} }$ and $prefix\_value \in R^ {L\times B\times {N_h}\times {d_h}}$. Where $L$ is the data length, $B$ denotes batch size, $N_h$ denotes the number of attention heads, and $d_h$ is the dimension of each head.

In the process of forward operation, when the data passes through each LLM-Block, the prefix is spliced with the frozen key and value in the model to form a new key' and value', which are denoted as $K^{'}$ and $V^{'}$, respectively, with the original query parameter $\left(Q \right)$ of the model to compute the attention score of the current data as well as the hidden state. Taking the i-th layer LLM-Block as an example, the computation process of p-tuning v2 is shown below:

\begin{align*}
&{key_i}^{'}: {K_i}^{'}=[{prefix\_key}_i:{key}_i] \\
&{{value}_i}^{'}: {V_i}^{'}=[{prefix\_value}_i:{value}_i] \\
&Attention\ score:S_i^\prime=softmax(\frac{Q_i{{K_i}^\prime}^T}{\sqrt{d_h}}) \\
&{hidden\_state}_i=FFN(S_i^\prime{V_i}^\prime)
\end{align*}

\section{Dataset}
\label{sec:datasets detail}
Table~\ref{tb:superglueTask} shows the cloze questions and answers for SuperGLUE tasks, and the detailed corresponding description of SuperGLUE benchmark are as below: 
\begin{itemize}
\item ReCoRD(Reading Comprehension with Commonsense Reasoning and Disambiguation): In this task, models are required to answer questions by extracting information from a given passage, while also employing commonsense reasoning and resolving ambiguous pronouns.

\item COPA(Choice of Plausible Alternatives): This task assesses causal reasoning abilities by providing a premise and two alternative hypotheses, where the model must choose the correct causal relationship.

\item WSC(Winograd Schema Challenge): This task evaluates pronoun resolution and coreference resolution abilities, where the model must identify the correct referent for a pronoun in a given sentence.

\item RTE(Recognizing Textual Entailment): The task requires determining if one sentence entails, contradicts, or remains neutral with respect to another sentence.

\item BoolQ(Boolean Questions): Models must answer boolean questions, i.e., questions that require a yes or no answer, based on a given context.

\item WiC(Word-in-Context): In this task, models must determine if a word has the same sense in two different contexts, requiring fine-grained lexical semantics understanding.

\item CB(CommitmentBank): It is a famous corpus of short texts for textual entailment task, in which at least one sentence contains an embedded clause. 

\item MultiRC(Multiple-Choice Reading Comprehension): This task involves answering multiple-choice questions based on multiple passages, which tests the ability to comprehend complex documents.
\end{itemize}

\begin{table}[!t]
  \centering  
    \begin{adjustbox}{width=0.48\textwidth}
    \begin{tabular}{ccccc}
    \toprule
    Datasets & Average Period & Sequential & client-batch parallel & server-hierarchical \\
    \midrule
    \multirow{2}[1]{*}{COPA} & 50    & 85    & 85    & 85 \\
          & 100   & 85    & 85    & 85 \\
    \multirow{2}[0]{*}{WiC} & 50    & 69.1 & 66.6 & 68.2 \\
          & 100   & 69.0 & 65.5 & 67.2 \\
    \multirow{2}[0]{*}{RTE} & 50    & 80.1 & 80.1 & 78.3 \\
          & 100   & 79.8 & 79.4 & 77.6 \\
    \multirow{2}[0]{*}{BoolQ} & 50    & 81.6 & 79.9 & 81.0 \\
          & 100   & 81.9 & 80.5 & 81.3 \\
    \multirow{2}[0]{*}{MultiRC} & 50    & 79.3 & 76.2 & 77.5 \\
          & 100   & 77.5 & 76.6 & 77.1 \\
    \multirow{2}[0]{*}{CB} & 50    & 85.7 & 85.7 & 85.7 \\
          & 100   & 85.7 & 85.7 & 85.7 \\
    \multirow{2}[1]{*}{WSC} & 50    & 71.2 & 63.5 & 63.5 \\
          & 100   & 66.3 & 65.4 & 63.5 \\
    \bottomrule
    \end{tabular}%
    \end{adjustbox}
    \caption{\label{tb:average period} Impact of different average period}
  
\end{table}%

\begin{figure*}[!t]
    \centering
    \includegraphics[width=1.0\linewidth]{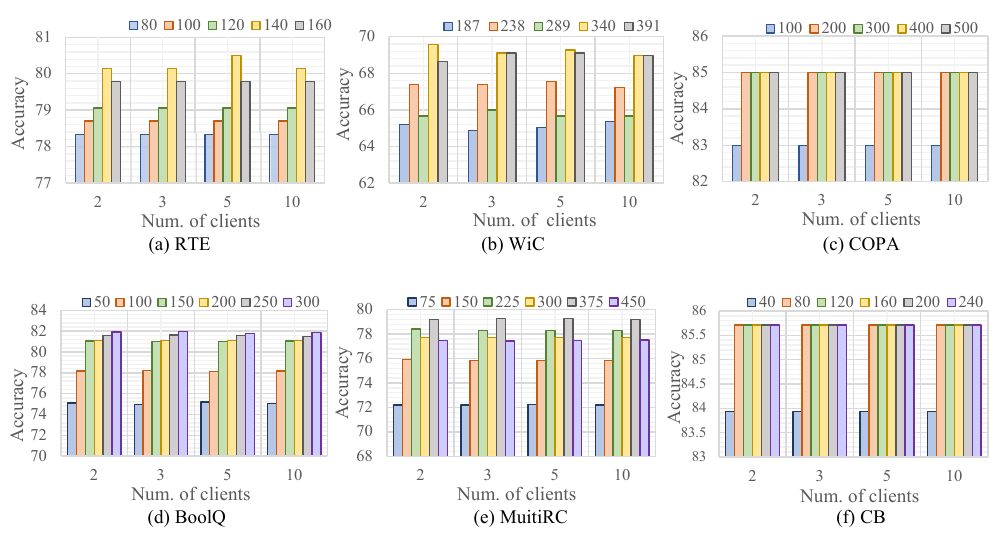}
    \caption{Comparison of model performance under serial training, where colors denote distinct training steps.}
    \label{fig:sequential test result}
\end{figure*}

\section{Impact of Average Period} \label{app:average}

For analyzing the effect of different averaging periods on the model performance, we tested the performance of FL-GLM with different averaging periods (50 step and 100 step).

The results are shown in Table ~\ref{tb:average period}, where the model with an average period of 100 steps slightly outperforms the model with an average period of 50 steps in the BoolQ task. However, in the WiC, RTE, and MultiRC tasks, better results are achieved with an average period of 50 steps. In the COPA and CB tasks, the averaging period has no effect on performance. The most noticeable difference occurs in the WSC task, with scores of 71.2 and 66.3 for an average period of 50 steps and 100 steps, respectively, for serial training, 63.5 and 65.4 for client-batch parallel, and flat accuracy scores for server-hierarchical. Among all the evaluation tasks, the WSC task has the highest sensitivity to the average period, but the average training period has little effect on the overall performance of the FL-GLM model with the same training strategy.

\begin{figure*}[!t]
    \centering
    \includegraphics[width=1.0\linewidth]{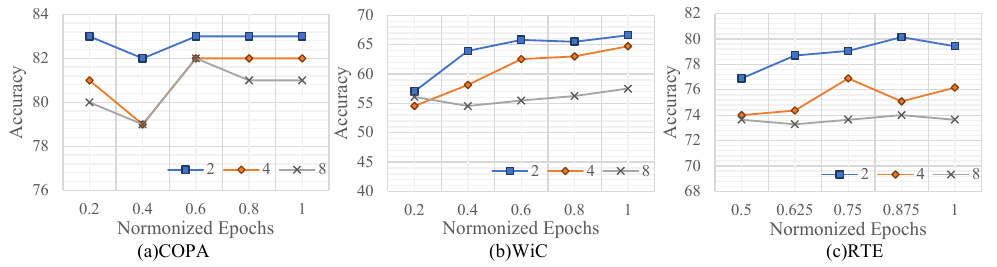}
    \caption{Comparison of the accuracy curves under varying numbers of clients using a client-batch parallel training.
    }
    \label{fig:batch parallel test result}
\end{figure*}

\begin{figure*}[!t]
    \centering
    \includegraphics[width=1.0\linewidth]{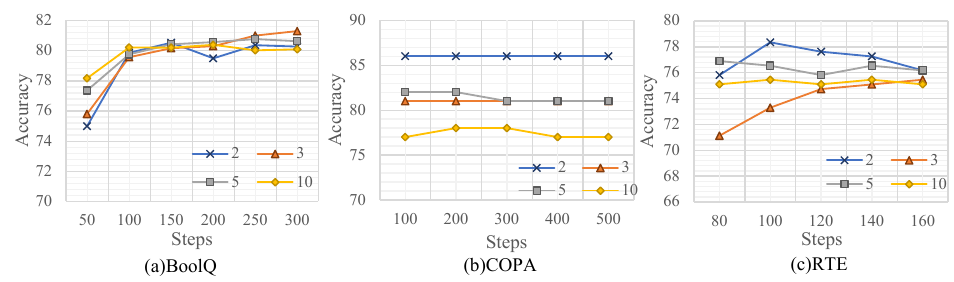}
    \caption{Comparison of the accuracy curves under varying numbers of clients using a server-hierarchical training.
    }
    \label{fig:server parallel test result}
\end{figure*}

\section{Impact of Participants} \label{app:part}
In this section, we test the three training strategies with different numbers of clients by calculating the accuracy scores of FL-GLM on different datasets.~\footnote{In the client-batch parallel test, in order to mitigate the effect of overfitting, the datasets are trained with the same number of training epochs for different numbers of clients, and normalization is used to enhance the visibility of the results.}The sequential test uses RTE, WiC, COPA, BoolQ, MultiRC and CB datasets, while the client-batch parallel test uses RTE, WiC, COPA datasets, and the server-hierarchical test uses BoolQ, COPA and RTE datasets, and the hyperparameters such as learning rate are kept consistent.


When using serial training strategy, the impact of increasing the number of clients is minimal, as shown in Figure \ref{fig:sequential test result}. This is because the majority of parameters are trained on the server, making the number of clients insignificant in server-side parameter training. 

When training in parallel, the accuracy score of FL-GLM decreases slightly as the number of clients increases, which is more obvious on datasets with smaller data volumes.
For client-batch parallel training, as shown in Figure~\ref{fig:batch parallel test result}, the accuracy score decreases with the increase in the number of clients due to the increase in the batch size, the frequency of model parameter updating decreases, and the server-side model is easy to converge to the saddle point.
For hierarchical-server parallel, as shown in Figure~\ref{fig:server parallel test result}, the increase in the number of clients makes the amount of data for a single client smaller, so the more the number of clients, the more obvious the overfitting phenomenon is.




\end{document}